\documentclass[letterpaper, 12pt]{article}
\usepackage{jheppub}

\usepackage{graphicx}
\usepackage{bbm}
\usepackage{float}
\usepackage{amssymb}
\usepackage{amsthm}
\usepackage{multirow}
\usepackage{physics}
\usepackage{afterpage}
\usepackage{rotating}
\usepackage{upgreek}
\usepackage{amsmath}
\usepackage{mathtools}
\usepackage{epstopdf}
\usepackage{mathrsfs}
\usepackage{scalefnt}
\usepackage[labelformat=simple]{subcaption}
\usepackage{calligra}
\usepackage[T1]{fontenc}  

\begin{document}

\begin{titlepage}

\setcounter{page}{1} \baselineskip=15.5pt \thispagestyle{empty}
{\flushright {ITP-CAS-25-362}\\}
		
\bigskip\
		
\vspace{2.1cm}
\begin{center}
{\LARGE \bfseries Toward the Effective Light and Heavy QCD\vspace{0.24cm}\\ Axion Scenarios}
\end{center}
\vspace{0.15cm}
			
\begin{center}
{\fontsize{14}{30}\selectfont Hai-Jun Li$^{a,b}$}
\end{center}
\begin{center}
\vspace{0.25 cm}
\textsl{$^a$Institute of Theoretical Physics, Chinese Academy of Sciences, Beijing 100190, China}\\
\textsl{$^b$International Centre for Theoretical Physics Asia-Pacific, Beijing 100190, China}\\

\vspace{-0.1 cm}				
\begin{center}
{E-mail: \textcolor{blue}{\tt {lihaijun@itp.ac.cn}}}
\end{center}	
\end{center}
\vspace{0.6cm}
\noindent

In this work, we investigate the effective parameter space associated with the axion mass and the axion decay constant in both the light and heavy QCD axion scenarios.
We initiate our discussion by considering the simplest case of two axions, quantitatively analyzing the parameter space in these two distinct scenarios.
We find that the axion mass ratios exhibit a high degree of similarity in these two situations.
In contrast, the ratios of axion decay constants display a complete opposition.
Furthermore, we generalize our conclusions to encompass the case of multiple axions.
 		
\vspace{3.5cm}
			
\bigskip
\noindent\today
\end{titlepage}
			
\setcounter{tocdepth}{2}
			
 

\section{Introduction}

It has long been a well-established fact that a plenitude of axions can have their origins in higher-dimensional gauge fields \cite{Witten:1984dg, Green:1984sg, Svrcek:2006yi, Conlon:2006tq, Reece:2025thc}.
We anticipate that among these axions, there exists one that can offer a solution to the strong CP problem in the Standard Model (SM) \cite{Peccei:1977hh, Peccei:1977ur, Weinberg:1977ma, Wilczek:1977pj}, with the rest being ultra-light axion-like particles (ALPs).
In particular, the type IIB string axiverse encompasses one QCD axion and many ALPs \cite{Arvanitaki:2009fg, Cicoli:2012sz, Broeckel:2021dpz}.
Moreover, these axions can also serve as dark matter candidates \cite{Preskill:1982cy, Abbott:1982af, Dine:1982ah}.

Axion mixing within the axiverse has recently garnered considerable attention, as evidenced by a range of studies \cite{Hill:1988bu, Daido:2015cba, Li:2023uvt, Cyncynates:2023esj, Kitajima:2014xla, Ho:2018qur, Li:2023xkn, Murai:2024nsp, Li:2024jko, Li:2025cep, Murai:2025wbg}.  
In particular, ref.~\cite{Li:2025cep} investigated axion mass mixing in the axiverse and employed a bottom-up approach to identify the conditions under which an axion model can achieve maximal mixing --- a scenario where the effective mixing degree reaches its peak.  
In the context of multiple axions, consisting of one QCD axion and many ALPs, maximal mixing occurs under specific conditions: when the masses of all ALPs are below the zero-temperature mass of the QCD axion, and when the decay constants of all ALPs are either uniformly smaller or uniformly larger than that of the QCD axion.  
Furthermore, these two situations --- smaller versus larger decay constants --- can be categorized as the light QCD axion scenario \cite{Daido:2015cba, Li:2023uvt} and the heavy QCD axion scenario \cite{Cyncynates:2023esj}, respectively.  
Each of these distinct scenarios brings about unique cosmological implications, with the most direct effect being the impact on axion energy density: in the light and heavy axion scenarios, the QCD axion energy density can be significantly suppressed and enhanced, respectively.
However, previous discussions regarding the relationships between axion masses and axion decay constants have remained purely qualitative, lacking quantitative analysis.

In this paper, we delve into the exploration of the effective parameter space within both the light and heavy QCD axion scenarios.
First, we conduct a concise review of the light and heavy axion scenarios in the context of multiple axions. 
Subsequently, we perform an in-depth analysis of the effective parameter space associated with the axion mass ratio ($\zeta$) and the axion decay constant ratio ($\eta$) in these two scenarios.
We initiate our discussion with the most straightforward case of two axions. 
Here, we separately examine the parameter space in both the light axion and heavy axion scenarios. 
The distributions of the delta mass eigenvalue $\Delta m$ and the delta mass eigenvalue at level crossing $\Delta m_\times$ are presented.
For the validity of the light and heavy axion scenarios, it is imperative that the value of $\Delta m_\times$ be minimized as much as possible. 
Additionally, we present $\Delta m_\times$ and the normalized $\langle\Delta m_\times\rangle$ in the $\{\zeta, \eta\}$ parameter plane.
Our findings reveal that the axion mass ratios exhibit similarity in these two scenarios, while the ratios of axion decay constants are diametrically opposed.
Following this, we generalize our conclusions to the case of multiple axions. 

The rest of this paper is structured as follows.  
In section~\ref{sec_overview}, we briefly review the light and heavy QCD axion scenarios.
In section~\ref{sec_parameter_space}, we analyze the effective parameter space of the axion mass and decay constant in both the light and heavy axion scenarios, extending from the simplest two-axion case to the multi-axion context.
Finally, the conclusion is given in section~\ref{sec_Conclusion}.

\section{Overview of the light and heavy QCD axion scenarios}
\label{sec_overview}

In this section, we briefly review the light and heavy QCD axion scenarios.
In the context of multiple axions, one QCD axion ($a$) and $N$ ALPs ($A_i$), the low-energy effective mixing potential is given by
\begin{eqnarray}
\begin{aligned}
V_{\rm mix}&=m_a^2 f_a^2\left[1-\cos\left(n_{00}\dfrac{\phi}{f_a}+\sum_{j=1}^N n_{0j}\dfrac{\varphi_j}{f_{A_j}}+\delta_0\right)\right]\\
&+\sum_{i=1}^N m_{A_i}^2 f_{A_i}^2\left[1-\cos\left(n_{i0}\dfrac{\phi}{f_a}+\sum_{j=1}^N n_{ij}\dfrac{\varphi_j}{f_{A_j}}+\delta_i\right)\right]\, .
\end{aligned}
\end{eqnarray}
Here, $\phi$ and $\varphi_i$ are the  corresponding axion fields, $m_a$ and $m_{A_i}$ the axion masses, $f_a$ and $f_{A_i}$ the axion decay constants, $n_{ij}$ the domain wall numbers, and $\delta_i$ the constant phases.
The terms $\delta_i$ are set to zero, which can be seen as equivalent to requiring an independent solution to the strong CP problem.

To exhibit maximal mixing, two key conditions must be satisfied: first, the masses of all ALPs must differ and be smaller than the zero-temperature mass of the QCD axion; and second, the decay constants of all ALPs must all be smaller or all be larger than that of the QCD axion. 
The two different situations in the second condition --- smaller or larger decay constants --- correspond to the light QCD axion scenario and the heavy QCD axion scenario, respectively. 
In these two cases, the matrix of domain wall numbers $n_{ij}$ can be expressed as
\begin{eqnarray}
\mathfrak{n}_{\mathfrak{ij}}=
\left(
\begin{array}{ccccc}
1  & ~ 0 & ~ 0 &~\cdots & ~ 0\\
1  & ~ 1 & ~ 0 &~\cdots & ~ 0\\
1  & ~ 0 & ~ 1 &~\cdots & ~ 0\\
\vdots  & ~\vdots &~\vdots & ~\ddots & ~ \vdots\\
1  & ~ 0 & ~ 0 & ~ \cdots & ~ 1
\end{array}
\right)\, , \quad
\mathfrak{n}_{\mathfrak{ij}}=
\left(
\begin{array}{ccccc}
1  & ~ 1 & ~ 1 &~\cdots & ~ 1\\
0  & ~ 1 & ~ 0 &~\cdots & ~ 0\\
0  & ~ 0 & ~ 1 &~\cdots & ~ 0\\
\vdots  & ~\vdots &~\vdots & ~\ddots & ~ \vdots\\
0  & ~ 0 & ~ 0 & ~ \cdots & ~ 1
\end{array}
\right)\, ,
\label{nij_max}
\end{eqnarray}
respectively.
Here $\mathfrak{n}_{\mathfrak{ij}}$ is a $(N+1)\times(N+1)$ matrix with indices starting from $\mathfrak{i}=0$ and $\mathfrak{j}=0$.
Further details on the dynamics and cosmological implications of maximal axion mixing are provided in ref.~\cite{Li:2025cep}. 

However, in previous literature, the discussions regarding the relationships between axion masses and between decay constants have been merely qualitative, lacking quantitative analysis. 
Therefore, in this work, we will conduct a quantitative analysis to determine the parameter space in both the light and heavy axion scenarios.

\section{Effective parameter ($\zeta, \eta$) space for light and heavy axions}
\label{sec_parameter_space}

In this section, we analyze the effective parameter space of the axion mass and decay constant in both the light and heavy QCD axion scenarios. 
For simplicity, we first take the simplest two-axion scenario as an example, but later the conclusions can be generalized to scenarios involving multiple axions.
 
The simplest scenario is to consider the two-axion model, which has been extensively studied in the literature.
Consider the QCD axion field $\phi$ and one ALP field $\varphi$. 
The mixing potential is given by
\begin{eqnarray}
\begin{aligned}
V_{\rm mix}&=m_a^2 f_a^2\left[1-\cos\left(n_{00}\dfrac{\phi}{f_a}+n_{01}\dfrac{\varphi}{f_A}\right)\right]+m_A^2 f_A^2\left[1-\cos\left(n_{10}\dfrac{\phi}{f_a}+ n_{11}\dfrac{\varphi}{f_A}\right)\right]\, .
\end{aligned}
\end{eqnarray}
For the convenience of subsequent discussions, we define two key quantities --- the axion mass ratio and the axion decay constant ratio:
\begin{eqnarray}
\zeta \equiv \dfrac{m_A}{m_{a,0}}\, ,\quad \eta \equiv \dfrac{f_A}{f_a}\, .
\end{eqnarray} 
Here the ALP is the most straightforward single-field model with the constant mass $m_A$. 
The term ${m_{a,0}}$ is the zero-temperature mass of the QCD axion \cite{GrillidiCortona:2015jxo}
\begin{eqnarray}
m_{a,0}=\dfrac{m_\pi f_\pi}{f_a}\dfrac{\sqrt{m_u/m_d}}{1+m_u/m_d}\simeq 5.70(7)\,{\mu \rm eV}\left(\dfrac{f_a}{10^{12}\,{\rm GeV}}\right)^{-1}\, ,
\label{ma0}
\end{eqnarray}
where $m_\pi$ and $f_\pi$ are the mass and decay constant of the pion, $m_u$ and $m_d$ are the up and down quark masses, respectively. 
Notice that when the temperature exceeds the critical temperature of the QCD phase transition $T_{\rm QCD}\simeq 150\, \rm MeV$, the QCD axion mass is temperature-dependent, $m_a\simeq m_{a,0}(T/T_{\rm QCD})^{-4.08}$, where 4.08 is an index derived from the dilute instanton gas approximation \cite{Borsanyi:2016ksw}.
In our following discussion, we take the QCD axion decay constant $f_a=10^{12}\,{\rm GeV}$ as a typical value.

\subsection{Light QCD axion scenario}

In this subsection, we discuss the light QCD axion scenario.
In the context of light axion scenario with two axions, the $2\times 2$ matrix of domain wall numbers in eq.~\eqref{nij_max} should be taken as 
\begin{eqnarray}
\mathfrak{n}_{2\times2}=
\left(
\begin{array}{cc}
1  & ~ 0\\
1  & ~ 1
\end{array}
\right)\, .
\end{eqnarray}
Assuming that the oscillation amplitudes of axion fields are substantially smaller than their corresponding decay constants, the mass mixing matrix is given by
\begin{eqnarray}
\mathbf{M}^2&=&
\left(\begin{array}{cc}
m_a^2+\dfrac{m_A^2 f_A^2}{f_a^2}  & ~ \dfrac{m_A^2 f_A}{f_a}\\
\dfrac{m_A^2 f_A}{f_a} & ~ m_A^2
\end{array}\right)
=m_A^2
\left(\begin{array}{cc}
\left(\dfrac{f_A}{f_a}\right)^2  & ~ \dfrac{f_A}{f_a}\\
\dfrac{f_A}{f_a} & ~ 1
\end{array}\right)
+
\left(\begin{array}{cc}
m_a^2 & ~0\\
0 & ~0
\end{array}\right)\\
&=&m_A^2
\left(\begin{array}{cc}
\eta^2  & ~\eta\\
\eta & ~ 1
\end{array}\right)
+
\left(\begin{array}{cc}
m_a^2 & ~0\\
0 & ~0
\end{array}\right)
\, ,
\end{eqnarray}
and the light and heavy mass eigenvalues are given by
\begin{eqnarray}
m_{h,l}^2=\dfrac{1}{2}\left[m_a^2+m_A^2+m_A^2\eta^2\pm \left(-4 m_a^2 m_A^2+\left(m_a^2+m_A^2+m_A^2\eta^2\right)^2\right)^{1/2} \right]\, .
\end{eqnarray}  
Here the light and heavy mass eigenvalues exhibit a unique level crossing phenomenon.
For our purposes, we define the delta mass eigenvalue 
\begin{eqnarray}
\Delta m \equiv m_h-m_l\, ,
\end{eqnarray}
and the delta mass eigenvalue at level crossing 
\begin{eqnarray}
\Delta m_\times \equiv \left(m_h-m_l\right)\big|_{T=T_\times}\, ,
\end{eqnarray}
where the level crossing temperature $T_\times$ is given by
\begin{eqnarray}
T_\times= T_{\rm QCD} \left(\zeta^2\left(1-\eta^2\right)\right)^{-\frac{1}{8.16}} \, .   
\label{Tx_light}
\end{eqnarray}
Notice that the level crossing temperature is obtained by solving the differential equation ${\rm d}(m_h^2-m_l^2)/{\rm d}T=0$, where the temperature range is above $T_{\rm QCD}$.
From eq.~\eqref{Tx_light}, we can obtain the relation $|\eta|<1$.  

\begin{figure}[t]
\centering
\includegraphics[width=0.49\textwidth]{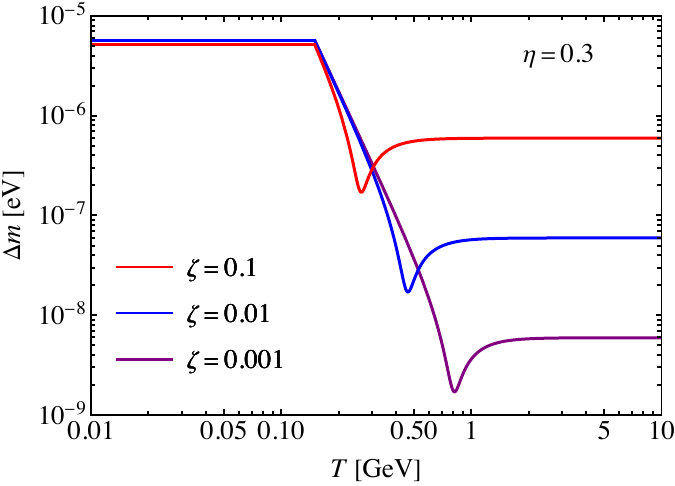}~\includegraphics[width=0.49\textwidth]{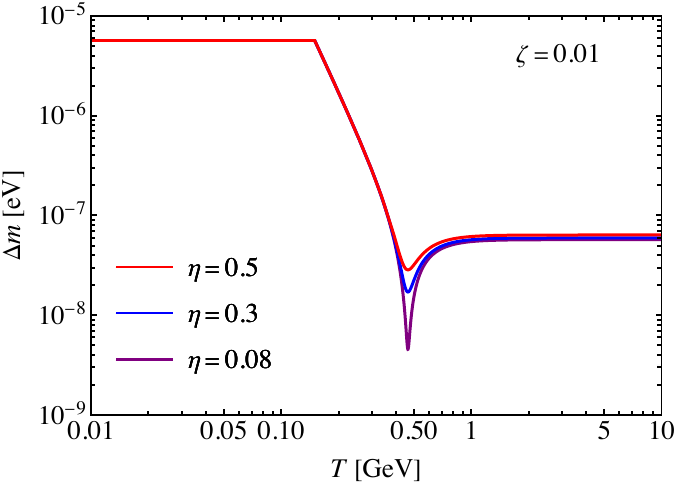}
\caption{{\bf Left panel}: The distribution of the delta mass eigenvalue $\Delta m$ with different values of the parameter $\zeta$.
Here we set $\eta=0.3$.
The red, blue, and purple solid lines represent $\zeta=0.1$, 0.01, and 0.001, respectively.
{\bf Right panel}: The distribution of the delta mass eigenvalue $\Delta m$ with different values of the parameter $\eta$.
Here we set $\zeta=0.01$.
The red, blue, and purple solid lines represent $\eta=0.5$, 0.3, and 0.08, respectively.
In these panels, we set $f_a=10^{12}\,{\rm GeV}$.}
\label{fig_deltam}
\end{figure} 

In figure~\ref{fig_deltam} (left panel), we present the distribution of the delta mass eigenvalue $\Delta m$ with different values of the mass parameter $\zeta$.
Here we fix the parameter $\eta$ and set $\eta=0.3$.
The red, blue, and purple solid lines represent $\zeta=0.1$, 0.01, and 0.001, respectively.
At high temperatures, the value of $\Delta m$ remains essentially constant because the corresponding $m_l$ value is very small at this time. 
As the temperature approaches $T_\times$, $\Delta m$ reaches its minimum value and then increases.
When the temperature is below $T_{\rm QCD}$, $\Delta m$ remains constant.
For different values of $\zeta$, $\Delta m$ exhibits significant differences at high temperatures, while the differences are almost negligible at low temperatures.
This is because at high temperatures, the values of $m_h$ corresponding to different $\zeta$ values vary significantly, whereas at low temperatures, $m_h$ is approximately equal to the zero-temperature mass of the QCD axion.
In the right panel of figure~\ref{fig_deltam}, we also present the distribution of the delta mass eigenvalue $\Delta m$ with different values of the decay constant parameter $\eta$.
Here we fix the parameter $\zeta$ and set $\zeta=0.01$.
The red, blue, and purple solid lines represent $\eta=0.5$, 0.3, and 0.08, respectively. 
In this case, for different values of $\eta$, the main differences occur at the level crossing temperature $T_\times$. 
The smaller the value of $\eta$, the smaller the corresponding value of the delta mass eigenvalue $\Delta m$, $\rm i.e.$, the value of $\Delta m_\times$.

\begin{figure}[t]
\centering
\includegraphics[width=0.49\textwidth]{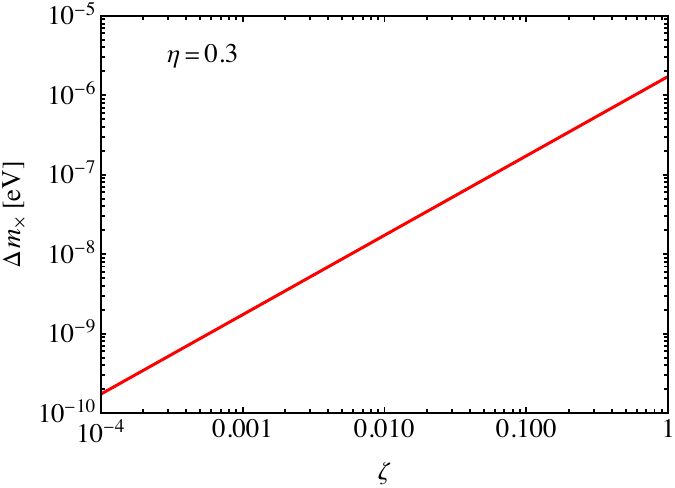}~\includegraphics[width=0.49\textwidth]{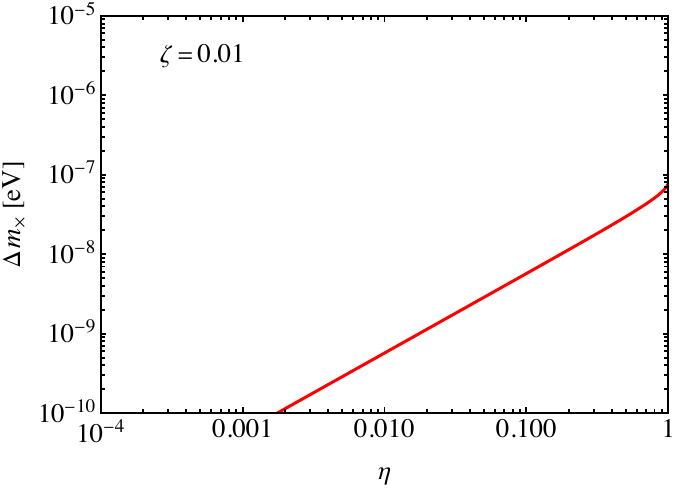}
\caption{{\bf Left panel}: The delta mass eigenvalue at level crossing $\Delta m_\times$ as a function of the parameter $\zeta$.
Here we set $\eta=0.3$.
{\bf Right panel}: The delta mass eigenvalue at level crossing $\Delta m_\times$ as a function of the parameter $\eta$.
Here we set $\zeta=0.01$.}
\label{fig_deltamx}
\end{figure}

To gain a better understanding of the variation in $\Delta m$ at $T_\times$, we present the delta mass eigenvalue at level crossing $\Delta m_\times$ as a function of the parameters $\zeta$ and $\eta$ in the left and right panels of figure~\ref{fig_deltamx}, respectively.
We find that the value of $\Delta m_\times$ decreases as both $\zeta$ and $\eta$ decrease. 
In this work, $\Delta m_\times$ plays a crucial role.
When $\Delta m_\times$ is significantly large, it implies a substantial discrepancy between the mass eigenvalues and the axion mass before mixing. 
Under such circumstances, we consider the light QCD axion scenario to be no longer valid.
Therefore, for the light axion scenario discussed here to be valid, the value of $\Delta m_\times$ should be as small as possible.\footnote{Notice that this condition is equally valid for the heavy axion scenario, as discussed later.}
Next, we will discuss the effective ranges for the parameters $\zeta$ and $\eta$.

\begin{figure}[t]
\centering
\includegraphics[width=0.5\textwidth]{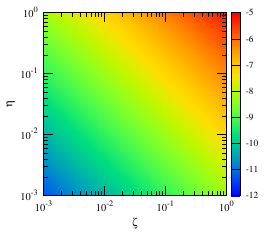}\includegraphics[width=0.5\textwidth]{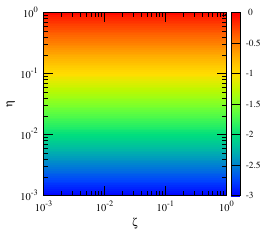}
\caption{{\bf Left panel}: The distribution of the delta mass eigenvalue at level crossing $\log_{10}(\Delta m_\times)$ in the $\{\zeta, \eta\}$ parameter plane.
{\bf Right panel}: The distribution of the normalized delta mass eigenvalue at level crossing $\log_{10}(\langle\Delta m_\times\rangle)$ in the $\{\zeta, \eta\}$ parameter plane.
Notice that in these panels $\zeta\neq 1$ and $\eta\neq 1$.}
\label{fig_deltamx_contour}
\end{figure}

More precisely, in figure~\ref{fig_deltamx_contour} (left panel), we present the distribution of the delta mass eigenvalue at level crossing $\Delta m_\times$ in the $\{\zeta, \eta\}$ parameter plane.
The $z$-axis represents the magnitude of $\log_{10}(\Delta m_\times)$.
Here, we also observe that to minimize $\Delta m_\times$, both parameters $\zeta$ and $\eta$ must take smaller values, $\rm i.e.$, $\zeta\ll 1$ and $\eta\ll 1$.
However, when $\eta$ is fixed, we find that the axion evolution at the level crossing remains similar despite variations in $\Delta m_\times$; see also figure~\ref{fig_deltam} (left panel).
Therefore, it is necessary to define the normalized delta mass eigenvalue at level crossing 
\begin{eqnarray}
\langle\Delta m_\times\rangle \equiv \dfrac{\left(m_h-m_l\right)\big|_{T=T_\times}}{m_h\big|_{T\to +\infty}}\, .
\end{eqnarray}
Here, the denominator represents the heavy mass eigenvalue at high temperatures.
In the right panel of figure~\ref{fig_deltamx_contour}, we present the distribution of the normalized delta mass eigenvalue at level crossing $\langle\Delta m_\times\rangle$ in the $\{\zeta, \eta\}$ parameter plane.
Here the $z$-axis represents the magnitude of $\log_{10}(\langle\Delta m_\times\rangle)$.
We find that the normalized value of $\langle\Delta m_\times\rangle$ remains independent of variations in the axion mass ratio $\zeta$.

\begin{figure}[t]
\centering
\includegraphics[width=0.49\textwidth]{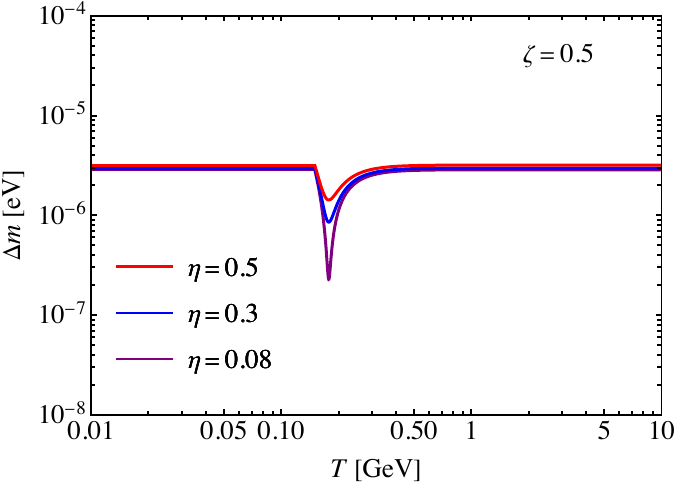}~\includegraphics[width=0.49\textwidth]{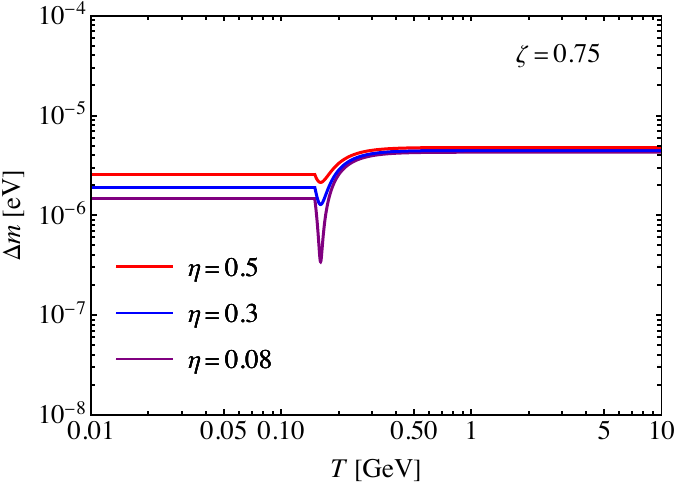}
\caption{Same as the right panel of figure~\ref{fig_deltam} but for larger values of the parameter $\zeta$.
{\bf Left panel}: We have $\Delta m\big|_{T\to 0}=\Delta m\big|_{T\to +\infty}$ with $\zeta=0.5$.
{\bf Right panel}: We have $\Delta m\big|_{T\to 0}<\Delta m\big|_{T\to +\infty}$ with $\zeta=0.75$.
The red, blue, and purple solid lines represent $\eta=0.5$, 0.3, and 0.08, respectively.}
\label{fig_deltam_add}
\end{figure}

Nevertheless, we argue that the parameter $\zeta$ should satisfy a condition where $\Delta m$ exhibits higher values at low temperatures compared to high temperatures,
\begin{eqnarray}
\Delta m\big|_{T\to 0}>\Delta m\big|_{T\to +\infty}\, .
\label{zeta_uneq}
\end{eqnarray}
This is because, for moderately large values of $\eta$, excessive $\zeta$ values disrupt the effective level crossing, rendering the light axion scenario discussed here invalid.
By solving
\begin{eqnarray}
\left(m_h-m_l\right)\big|_{T\to 0}=\left(m_h-m_l\right)\big|_{T\to +\infty}\, ,
\end{eqnarray}
we obtain
\begin{eqnarray}
\zeta\simeq\dfrac{1}{2}\, .
\end{eqnarray}
As an example, in the left and right panels of figure~\ref{fig_deltam_add}, we present the distribution of the delta mass eigenvalue $\Delta m$ for larger values of $\zeta=0.5$ and 0.75, respectively.
Particularly, the red line (with $\zeta=0.75$ and $\eta=0.5$) in the right panel shows that the value of $\Delta m_\times$ is almost equal to $\Delta m\big|_{T\to 0}$, implying that the effective level crossing can be disrupted in this case.
In conclusion, our analysis shows that the viability of the light QCD axion scenario requires the parameter space of the axion mass ratio and decay constant ratio to satisfy the following relation
\begin{eqnarray}
0<\zeta\lesssim\dfrac{1}{2}\, ,\quad 0<\eta\ll 1\, .
\end{eqnarray}
Notice that it is hard to establish a strict upper bound for $\eta$.
However, we maintain that the light axion scenario is valid when $\eta$ is sufficiently small, specifically on the order of $\sim\mathcal{O}(0.1)$.
It is not that level crossing cannot occur when $\eta$ is greater than 0.1. 
What we mean is that when effective level crossing takes place, we need to ensure that the light axion scenario remains valid, which implies that the parameter $\eta$ should be as small as possible.

Next, we will use the same method to analyze the heavy axion scenario.

\subsection{Heavy QCD axion scenario}

In this subsection, we discuss the heavy QCD axion scenario with two axions, employing an analytical approach similar to that of the light QCD axion scenario in the previous subsection.
In this context, the $2\times 2$ matrix of domain wall numbers in eq.~\eqref{nij_max} should be taken as 
\begin{eqnarray}
\mathfrak{n}_{2\times2}=
\left(
\begin{array}{cc}
1  & ~ 1\\
0  & ~ 1
\end{array}
\right)\, .
\end{eqnarray}
The mass mixing matrix is given by
\begin{eqnarray}
\mathbf{M}^2&=&
\left(\begin{array}{cc}
m_a^2  & ~ \dfrac{m_a^2 f_a}{f_A}\\
\dfrac{m_a^2 f_a}{f_A} & ~ m_A^2+\dfrac{m_a^2 f_a^2}{f_A^2}
\end{array}\right)
=m_a^2
\left(\begin{array}{cc}
1  & ~ \dfrac{f_a}{f_A}\\
\dfrac{f_a}{f_A} & ~\left(\dfrac{f_a}{f_A}\right)^2
\end{array}\right)
+
\left(\begin{array}{cc}
0 & ~0\\
0 & ~m_A^2
\end{array}\right)\\
&=&m_a^2
\left(\begin{array}{cc}
1  & ~1/\eta\\
1/\eta & ~ 1/\eta^2
\end{array}\right)
+
\left(\begin{array}{cc}
0 & ~0\\
0 & ~m_A^2
\end{array}\right)
\, ,
\end{eqnarray}
and the light and heavy mass eigenvalues are given by
\begin{eqnarray}
m_{h,l}^2=\dfrac{1}{2\eta^2}\left[m_a^2+m_a^2\eta^2+m_A^2\eta^2\pm \left(-4 m_a^2 m_A^2\eta^4+\left(m_a^2+m_a^2\eta^2+m_A^2\eta^2\right)^2\right)^{1/2} \right]\, .~~~~
\end{eqnarray}  
Here we also define the delta mass eigenvalue $\Delta m$, the delta mass eigenvalue at level crossing $\Delta m_\times$, and the normalized delta mass eigenvalue at level crossing $\langle\Delta m_\times\rangle$ in the same way as in the previous subsection. 
In this case, the level crossing temperature is provided as
\begin{eqnarray}
T_\times= T_{\rm QCD} \left(\dfrac{\zeta^2 \eta^2\left(\eta^2-1\right)}{\left(\eta^2+1\right)^2}\right)^{-\frac{1}{8.16}} \, .    
\label{Tx_heavy}
\end{eqnarray}
From eq.~\eqref{Tx_heavy}, we can obtain the relation $|\eta|>1$.  

\begin{figure}[t]
\centering
\includegraphics[width=0.49\textwidth]{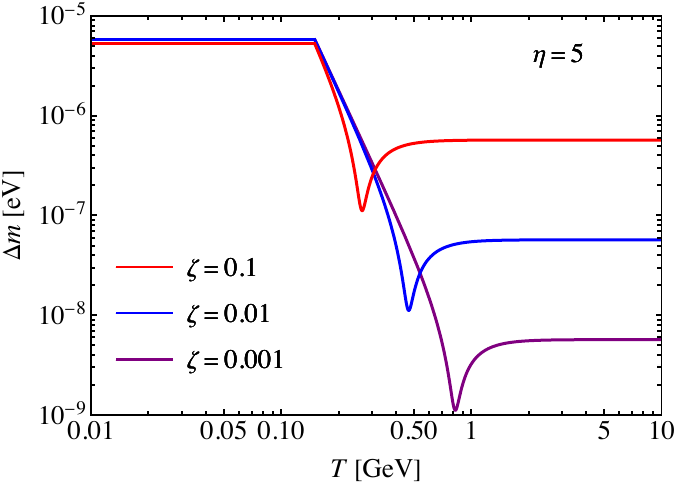}~\includegraphics[width=0.49\textwidth]{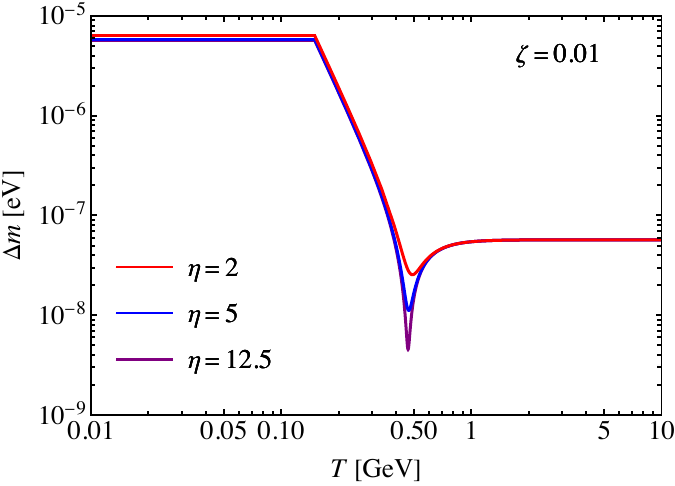}
\caption{Same as figure~\ref{fig_deltam} but for the heavy QCD axion scenario.
{\bf Left panel}: The distribution of the delta mass eigenvalue $\Delta m$ with different values of the parameter $\zeta$.
Here we set $\eta=5$.
The red, blue, and purple solid lines represent $\zeta=0.1$, 0.01, and 0.001, respectively.
{\bf Right panel}: The distribution of the delta mass eigenvalue $\Delta m$ with different values of the parameter $\eta$.
Here we set $\zeta=0.01$.
The red, blue, and purple solid lines represent $\eta=2$, 5, and 12.5, respectively.
We set $f_a=10^{12}\,{\rm GeV}$.}
\label{fig_deltam_heavy}
\end{figure}  
 
Since the discussion of the parameter $\zeta$ in the heavy axion scenario is similar to that in the light axion scenario, we will here focus our attention on the parameter $\eta$.
In figure~\ref{fig_deltam_heavy}, we present the distribution of the delta mass eigenvalue $\Delta m$ with different values of the parameters $\zeta$ and $\eta$.
In the left panel, we can observe that the evolution of $\Delta m$ is similar to that in figure~\ref{fig_deltam}. 
It is important to note that here we set $\eta>1$ and fix it to $\eta=5$.
In the right panel of figure~\ref{fig_deltam_heavy}, we present the distribution of $\Delta m$ for different values of $\eta$. 
Here we also fix the parameter $\zeta$ and set $\zeta=0.01$.
The red, blue, and purple solid lines represent $\eta = 2$, 5, and 12.5, respectively. 
For different values of $\eta$, the main differences occur at the level crossing temperature $T_\times$. 
However, the larger the value of $\eta$, the smaller the corresponding value of $\Delta m_\times$.
This is the main difference between the light and heavy axion scenarios concerning the parameter $\eta$.

\begin{figure}[t]
\centering
\includegraphics[width=0.5\textwidth]{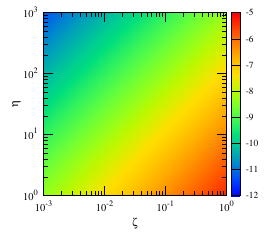}\includegraphics[width=0.5\textwidth]{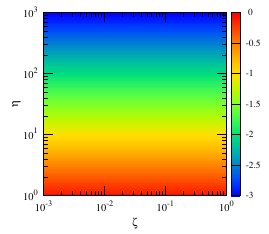}
\caption{Same as figure~\ref{fig_deltamx_contour} but for the heavy axion scenario.
{\bf Left panel}: The distribution of the delta mass eigenvalue at level crossing $\log_{10}(\Delta m_\times)$ in the $\{\zeta, \eta\}$ parameter plane.
{\bf Right panel}: The distribution of the normalized delta mass eigenvalue at level crossing $\log_{10}(\langle\Delta m_\times\rangle)$ in the $\{\zeta, \eta\}$ parameter plane.
Here $\zeta\neq 1$ and $\eta\neq 1$.}
\label{fig_deltamx_contour_heavy}
\end{figure} 

Furthermore, in figure~\ref{fig_deltamx_contour_heavy} (left panel), we present the distribution of the delta mass eigenvalue at level crossing $\log_{10}(\Delta m_\times)$ in the $\{\zeta, \eta\}$ parameter plane.
We find that to minimize $\Delta m_\times$, the parameter $\zeta$ should take a smaller value, while the parameter $\eta$ should take a larger value, $\rm i.e.$, $\zeta\ll 1$ and $\eta\gg 1$.
In the right panel of figure~\ref{fig_deltamx_contour_heavy}, we also present the distribution of the normalized delta mass eigenvalue at level crossing $\log_{10}(\langle\Delta m_\times\rangle)$ in the $\{\zeta, \eta\}$ parameter plane.
As is discussed in the light axion scenario, here the normalized value of $\langle\Delta m_\times\rangle$ remains independent of variations in $\zeta$. 
In addition, by solving eq.~\eqref{zeta_uneq}, we can also obtain the constraint on the parameter $\zeta$ in the heavy axion scenario, namely $\zeta\lesssim 1/2$. 
In conclusion, the viability of the heavy QCD axion scenario requires the parameter space of the axion mass ratio and decay constant ratio to satisfy the following relation
\begin{eqnarray}
0<\zeta\lesssim\dfrac{1}{2}\, ,\quad \eta\gg 1\, .
\end{eqnarray}
It is equally difficult to set a strict lower bound for $\eta$ in this case, yet we assert that the heavy axion scenario remains valid when $\eta$ is sufficiently large, typically on the order of $\sim\mathcal{O}(10)$.
Here, level crossing can also occur when $\eta$ is less than 10, but $\eta$ should be as large as possible.

\subsection{Scenarios with multiple axions}
 
In the two preceding subsections, we discussed the simplest case of two axions.  
In this subsection, we extend our previous conclusions regarding the light and heavy QCD axion scenarios to encompass cases involving multiple axions.

Consider the scenario with one QCD axion ($a$) and $N$ ALPs ($A_i$). 
For the sake of convenience, we assume that the ALP masses have the sequence
\begin{eqnarray}
m_{A_1}<m_{A_2}<\cdots<m_{A_N}\, .
\end{eqnarray}
Note also that $m_{A_i}<m_{a,0},\, \forall i$. 
Then we define the axion mass ratios
\begin{eqnarray}
\zeta_i \equiv \dfrac{m_{A_i}}{m_{a,0}}\, , \quad \tilde\zeta_i \equiv \dfrac{m_{A_i}}{m_{A_{i+1}}}\, ,
\end{eqnarray}
and the axion decay constant ratio
\begin{eqnarray}
\eta_i \equiv \dfrac{f_{A_i}}{f_a}\, ,
\end{eqnarray} 
where ${m_{a,0}}$ and $f_a$ are the zero-temperature mass and the decay constant of the QCD axion.
In the previous sections, we considered the light and heavy QCD axion scenarios under the two-axion case and reached the conclusion that $\zeta\lesssim 1/2$. 
This is because, for moderately large values of $\eta$, excessive $\zeta$ values disrupt the effective level crossing, rendering the light or heavy axion scenario invalid. 
When considering multiple ALPs, by solving eq.~\eqref{zeta_uneq} in this context, we can also obtain the relationship between the ALP masses, namely
\begin{eqnarray}
\tilde\zeta_i \lesssim\dfrac{1}{2}\, .
\end{eqnarray}
Notice that this quantity represents the relation where two ALPs have adjacent masses.
Therefore, regardless of whether it is in the light or heavy QCD axion scenario, we conclude that
\begin{eqnarray}
\zeta_i\lesssim\left(\dfrac{1}{2}\right)^{N-i+1}\, .
\end{eqnarray}
When $N=1$, that is, in the two-axion scenario, we can also revert to our previous conclusion that $\zeta\lesssim 1/2$.  
On the other hand, since there are no restrictions on the decay constants of ALPs and they can be equal, the previous discussions on the axion decay constants in the two-axion case are equally applicable here. 
Finally, we conclude that for the multi-axion context, in the light QCD axion scenario
\begin{eqnarray}
0<\zeta_i\lesssim\left(\dfrac{1}{2}\right)^{N-i+1}\, ,\quad 0<\eta_i\ll 1\, ,
\end{eqnarray}
and in the heavy QCD axion scenario
\begin{eqnarray}
0<\zeta_i\lesssim\left(\dfrac{1}{2}\right)^{N-i+1}\, ,\quad \eta_i\gg 1\, .
\end{eqnarray}

\section{Conclusion}
\label{sec_Conclusion}

In summary, we have investigated the effective parameter space, denoted by ($\zeta, \eta$), which is linked to the axion mass and the axion decay constant, spanning both the light and heavy QCD axion scenarios.

To begin with, we conduct a concise review of the light and heavy axion scenarios within the broader context of multiple axions. 
Starting with the simplest case of two axions, one QCD axion and one ALP, we examine the axion mass ratio ($\zeta$) and the axion decay constant ratio ($\eta$) within these two distinct scenarios. 
We present the distributions of the delta mass eigenvalue $\Delta m$ and the delta mass eigenvalue at the level crossing $\Delta m_\times$, which are crucial for understanding the axion evolution. 
Notably, we emphasize that minimizing the value of $\Delta m_\times$ is of great importance for the validity of both light and heavy axion scenarios.
Furthermore, we present $\Delta m_\times$ and the normalized $\langle\Delta m_\times\rangle$ within the $\{\zeta, \eta\}$ parameter plane, providing a clear representation of these parameters.
Our findings reveal intriguing patterns: while the axion mass ratios exhibit similarities across the two scenarios, the ratios of axion decay constants display stark contrasts, highlighting the distinct nature of these scenarios.
Finally, we generalize our conclusions to encompass the more complicated case of multiple axions, thereby extending the applicability and relevance of our study. 
Overall, this work contributes valuable insights into the effective parameter space of the light and heavy QCD axion scenarios in the context of multiple axions.

\section*{Acknowledgments}

This work was partly supported by the Institute of Theoretical Physics, CAS, and partly supported by the International Centre for Theoretical Physics Asia-Pacific.



 
\bibliographystyle{JHEP}
\bibliography{references}

\end{document}